\documentclass[a4paper,11pt]{article}
\usepackage{pos}
\usepackage{aas_macros}

\title{Classification of Unidentified Extended LHAASO Sources based on their Gamma-Ray Morphology: Prospects for Future IACTs.}

\author*[a,b,c]{Alberto Bonollo}
\author[a,c]{Paolo Esposito}
\author[c]{Andrea Giuliani}
\author[c]{Silvia Crestan}
\author[c]{Giorgio Galanti}
\author[c]{Sandro Mereghetti}
\author[d,c]{Michela Rigoselli}

\affiliation[a]{Scuola Universitaria Superiore IUSS Pavia, \\
              Piazza della Vittoria 15, Pavia, Italy}

\affiliation[b]{Dipartimento di Fisica, Università degli Studi di Trento, \\
             Via Sommarive 14, Povo (TN), Italy}

\affiliation[c]{INAF -- Istituto di Astrofisica Spaziale e Fisica Cosmica (IASF) di Milano,\\
             Via A. Corti 12, I-20133 Milano, Italy}

\affiliation[d]{INAF -- Osservatorio Astronomico di Brera, \\
             Via Brera 28, I-20121 Milano, Italy}

\emailAdd{alberto.bonollo@inaf.it}

\abstract{While Supernova Remnants (SNRs) are widely considered the primary accelerators of cosmic rays (CRs) up to hundreds of TeV, they struggle to account for the CR flux at PeV energies, suggesting the existence of additional PeVatrons. Observations from LHAASO (Large High Altitude Air Shower Observatory) have identified several PeVatron candidates, including some SNRs, pulsar wind nebulae, TeV halos and young massive star clusters (YMSCs). These objects accelerate particles that interact with the surrounding interstellar medium and radiation fields, producing very-high-energy gamma rays (>100 TeV), a key signature of both leptonic and hadronic PeVatrons.

We simulate and model the emission of TeV halos and YMSCs, adopting radial emission profiles derived from observational data. Given the current angular resolution of gamma-ray instruments, these profiles often appear similar, making it challenging to distinguish between source classes. We explore how next-generation Imaging Atmospheric Cherenkov Telescopes (IACTs), namely the CTAO (Cherenkov Telescope Array Observatory) and the ASTRI Mini-Array (Astrofisica con Specchi a Tecnologia Replicante Italiana), can classify these sources based on their morphology. We test our classification methods, derived from the profile features of known sources, on simulated CTAO and ASTRI Mini-Array observations of unidentified extended sources from the first LHAASO catalog.

We present the results of our analysis to highlight the potential of future IACT observations in identifying the nature of extended gamma-ray sources, refining PeVatron candidate classifications, and improving our understanding of cosmic-ray accelerators.}

\ConferenceLogo{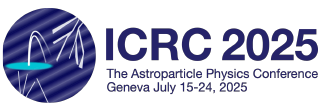}

\FullConference{39th International Cosmic Ray Conference (ICRC2025)\\
 15–24 July 2025\\
Geneva, Switzerland\\}

\begin{document}
\maketitle

\section{Introduction}
The origin of Galactic ultra-high energy (UHE) cosmic rays (CRs) remains one of the most important unsolved problems in astroparticle physics. While there is a broad agreement that cosmic ray nuclei and protons observed below $\sim3\times10^{15}$ eV,   the so-called "knee" in the cosmic ray energy spectrum,  originate within the Milky Way, the detailed properties of their acceleration sites are controversial. Theoretical investigations and observational limits, however, indicate that SNRs might face difficulties in accelerating particles to energies exceeding several hundreds TeV, i.e., roughly an order of magnitude below the PeV range. \citep{Gab}\\
In the past years, YMSCs have become a popular alternative and complementary targets for PeV cosmic ray production. YMSCs are formed by very crowded populations of massive stars, which give rise to powerful collective stellar winds that can develop into large-scale shock fronts, potentially accelerating particles to high enough energies. Under realistic astrophysical conditions--i.e. termination shock radius of $\sim$10 pc, magnetic field strengths of order 10 $\mu$G, and stellar wind velocities of the order of $\sim$1000 km/s--YMSCs can fulfill the requirements of the Hillas criterion, enabling the acceleration of particles to energies of $\sim$1 PeV.\citep{morlino, gabici} \\
While direction information is lost for charged cosmic rays due to deflections caused by Galactic magnetic fields, indirect information about their acceleration locations can be gained from observations of high-energy gamma rays. Gamma rays with energies above 100 TeV can be generated through collisions of such cosmic rays with ambient interstellar gas and radiation fields. These gamma-ray signals are the fingerprints of candidate PeVatrons—astrophysical accelerators that can accelerate particles up to PeV energies—regardless of whether their character is hadronic or leptonic. The next generation of IACT telescopes will greatly enhance our ability to investigate such high-energy domains thanks to  their wide fields of view (approximately 10$^\circ$ diameter) and better angular resolution (a few arcminutes at 1 TeV)\citep{astri,ctao}. CTAO, together with the ASTRI Mini-Array, will be more capable than current facilities (e.g. HESS, HAWC) and will allow morphological studies in fine detail of extended gamma-ray sources \citep{Aharonian1}.
Significantly, YMSCs and TeV halos are also conveniently sized for ground-based Cherenkov telescope observations since the angular extent on the sky of their emission region is typically $\sim1^{\circ}$. This spatial resolvability renders them within the reach of the CTAO and ASTRI Mini-Array, enabling prospects for in-depth investigations into the structure and radiation properties of these potential CR accelerators.\\
In this work, we explore the detectability and resolvability potential of ASTRI Mini-Array and CTAO for gamma-ray emission from YMSCs. Considering an optimal sample of selected sources, we establish our extended sources classification criteria. We then test CTAO and ASTRI Mini-Array capabilities on a sample of unidentified sources taken from the first LHAASO Catalog.

\section{Sample}
\begin{table*}
\centering
\begin{tabular}{lcccccc}
\hline\hline
Cluster & Distance (kpc) & Age (Myr) & $L_{\text{kin}}$ (erg s$^{-1}$) & $v_{\text{w}}$ (km s$^{-1}$) & $R_{\text{b}}$ (pc) & $R_{\text{ts}}$ (pc) \\
\hline
Cygnus OB2   & 1.6 & 1--6   & $2\times10^{38}$ & 2500 & 97.1  & 15.2 \\
Westerlund 1 & 3.9 & 3--10  & $1\times10^{39}$ & 3000 & 159.3 & 25.2  \\
Danks 1      & 3.8 & 1.5    & $8\times10^{37}$ & 3000 & 53.4  & 8.0   \\
Danks 2      & 3.8 & 3      & $7\times10^{37}$ & 3000 & 78.8  & 10.1  \\
Markarian 50 & 3.4 & 7.5    & $9\times10^{36}$ & 3000 & 90.5  & 7.9  \\
\hline
\end{tabular}
\caption{Physical properties of the selected YMSCs, including their distances, ages, wind luminosities ($L_{\text{kin}}$), assumed wind velocities ($v_{\text{w}}$), and the resulting bubble ($R_{\text{b}}$) and termination shock ($R_{\text{ts}}$) radii, expressed in both parsecs and degrees. Distances and ages are from: Cygnus OB2 \citep{2021MNRAS.502.6080O,2019MNRAS.484.1838B,Cyg5myr,cygage,cygagelow,Cyg10myr}; Westerlund 1 \citep{Wd1,wd1age}; Danks 1/2 \citep{danks}; Markarian 50 \citep{markarian}. Wind luminosities are from \cite{celli}. Wind velocities are from \cite{gabici,morlino,Aharonian1}. $R_\text{b}$ and $R_\text{ts}$ computed using Eq.~\ref{eq:rbrts}.}
\label{tab:parametri_ymsc}
\end{table*}
We compare the expected emission morphology of YMSCs with that of TeV halos, which are extended gamma-ray sources surrounding pulsars. We specifically include halos around the Geminga and Monogem pulsars \citep{2017Sci...358..911A} in our comparison, as they share key features with YMSCs: Galactic location, emission in the TeV range, and angular extensions of several degrees. This comparison helps inform future classification schemes and optimize observational strategies with upcoming instruments. We selected a sample of five YMSCs and 2 TeV halos for this study. To simulate their emission, we matched each source to an instrument response function (IRF) according to its location. Cygnus OB2, Markarian 50, Geminga and Monogem, in the northern hemisphere, were simulated using the IRF of the ASTRI Mini-Array. The southern clusters—Westerlund 1, Danks 1 and 2—were modeled with the IRF of the CTAO Southern Array.\\
Cygnus OB2 is an OB association characterised by a high mass and total luminosity. It is also in spatial coincidence with LHAASO-detected PeV photons \citep{CygLHAASO}. Given the relatively dispersed nature of its massive stars, the merging of their winds is expected to occur at lower velocities, hence we assume a reduced wind speed of 2500 km s$^{-1}$ \citep{vieu_morlino}. Distance estimates range from 1.6 to 1.8 kpc; we adopt 1.6 kpc, while age estimates vary between 2–6 Myr and we assume 3 Myr \citep{Cyg5myr,Cyg10myr}.\\
Westerlund 1 is among the most luminous YMSCs in the Galaxy, already observed in TeV gamma rays with H.E.S.S. \citep{Wd1}. Its distance is estimated at 3.9 kpc. The cluster's age remains uncertain, with values spanning 3–10 Myr depending on the assumed stellar evolution models and the history of star formation bursts \citep{wd1age}. We adopt 4 Myr as a representative age.\\
Danks 1 and 2 are two compact YMSCs located close to each other and are considered among the most luminous TeV YMSC candidates \citep{mitchell}. Their relative isolation in the sky and comparable distances ($\sim$3.8 kpc) make them promising targets for observations. We adopt distances and ages from \cite{danks} and simulated them both individually and together given their small relative distance.\\
Markarian 50, while older (7.5 Myr), is still relevant due to its relatively high wind luminosity and favorable northern location. Its age suggests possible contributions from supernova remnants (SNRs), which may enhance the gamma-ray signal. However, its isolation and brightness make it an excellent candidate for northern hemisphere observations \citep{markarian}.\\
We model the stellar wind-driven bubbles surrounding each YMSC, computing both the bubble radius $R_{\text{b}}$ and termination shock radius $R_{\text{ts}}$ within it as follows \citep{Aharonian1,koo}:
\begin{equation}
\begin{split}
\label{eq:rbrts}
    R_{\mathrm{b}} &= 0.88\,L_{\mathrm{w}}^{0.2}\,\rho_{\mathrm{H}}^{-0.2} \,t_{\mathrm{age}}^{0.6}\,\mathrm{pc},\\
    R_{\mathrm{ts}} &= 0.92\,L_{\mathrm{w}}^{0.3} \, v_{\mathrm{w}}^{-0.5} \, \rho_{\mathrm{H}}^{-0.3}\, t_{\mathrm{age}}^{0.4}\,\mathrm{pc}.
\end{split}
\end{equation}
Here, $L_{\mathrm{w}}$ is the mechanical wind luminosity of the cluster, $v_{\mathrm{w}}$ the average wind speed, and $\rho_{\mathrm{H}} = n_{\mathrm{H}} m_{\mathrm{p}}$ the ISM density. We assume a uniform ISM density of $n_{\mathrm{H}} = 10$ cm$^{-3}$ and a mass-loss rate of $10^{-4}\, M_\odot$/yr, consistently with \cite{morlino}.

\section{Source Simulations and Classification Criteria}
\begin{figure}
\centering
\includegraphics[width=0.48\linewidth]{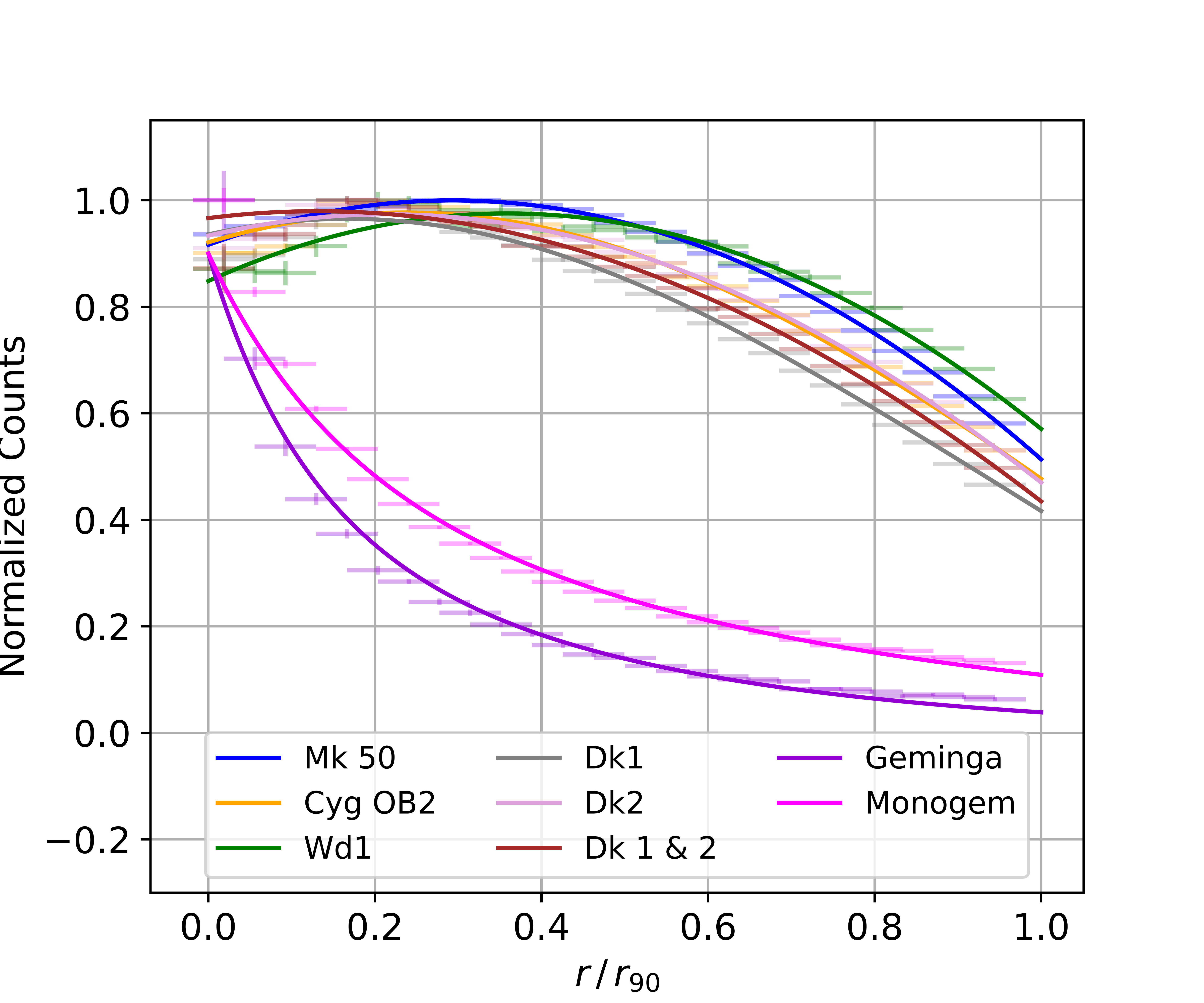}
\includegraphics[width=0.48\linewidth]{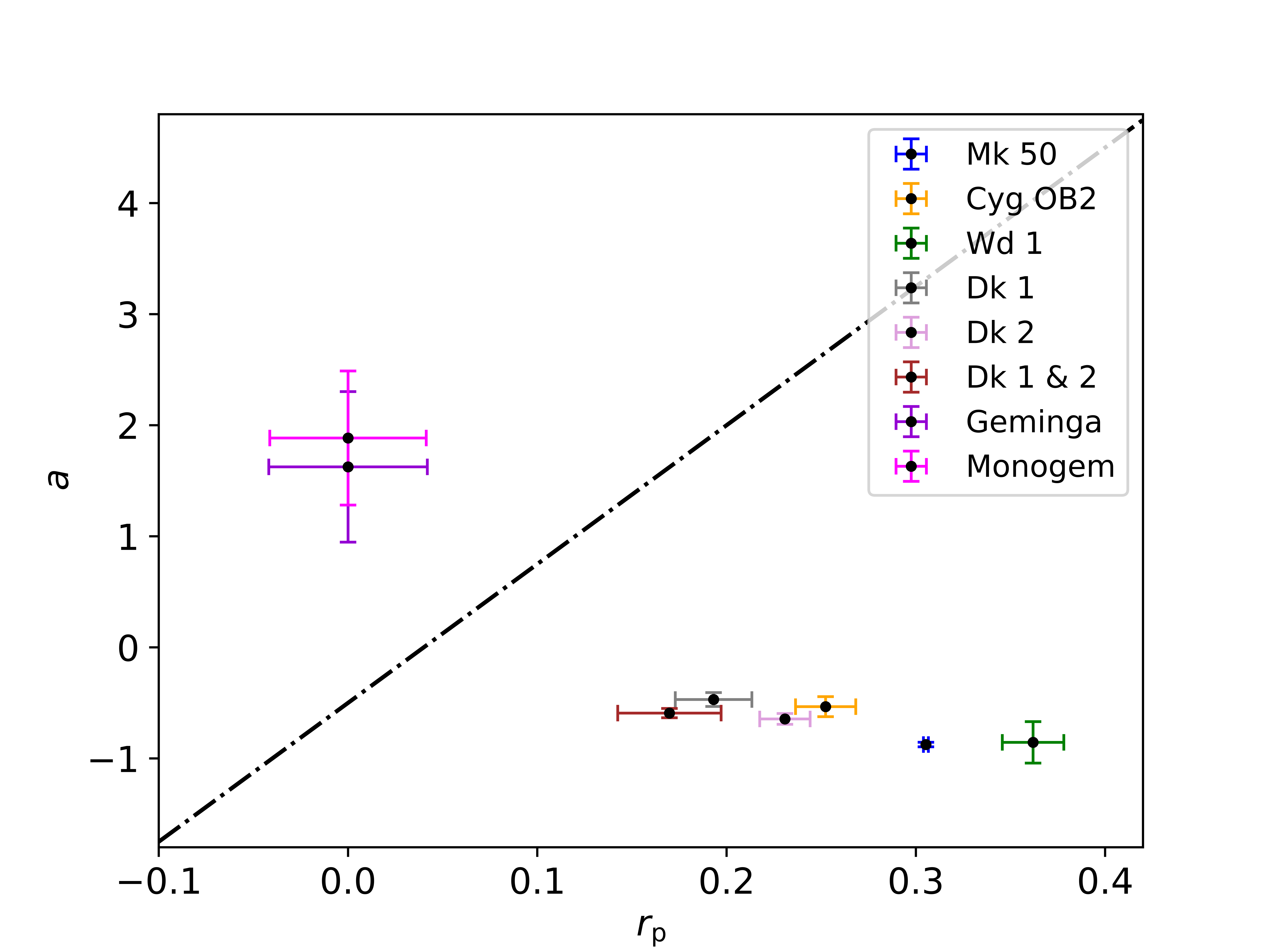}
\caption{\textit{Left}: Normalized radial gamma-ray profiles for all YMSCs and TeV halos. Each profile is scaled by its peak value for comparison. \textit{Right}: Anisotropy ($a$) vs. peak position ($r_p$) from modified Lorentzian fits. YMSCs cluster at lower $a$ and higher $r_p$, while TeV halos lie above the dot-dashed line.}
\label{fig:radial_profiles_spheric}
\end{figure}
We simulated gamma-ray emission in the 1–200 TeV range for all sources with the Gammapy software package \citep{gammapy:2023, gammapy_zenodo}. For TeV halos Geminga and Monogem, we utilized the 2D radiation profile of \cite{2017Sci...358..911A}, providing the energy- and angle-dependent flux
\begin{equation}
\frac{d^2 N}{dE d\Omega} = N_0 \left( \frac{E}{20~\mathrm{TeV}} \right)^{-\alpha} \frac{1.22}{\pi^{3/2} \theta_d(E)(\theta + 0.06\,\theta_d(E))} \exp\left( -\frac{\theta^2}{\theta_d(E)^2} \right),
\end{equation}
where $\theta_d$ is the diffusion angle. We employed a fine angular grid (2.5') and extrapolated the model to the entire energy range.\\
For YMSCs, we created 3D proton flux maps that depend on energy based on the hadronic model of \cite{morlino} under diffusive shock acceleration at the termination shock. Proton spectra were converted to gamma-ray templates via Naima libraries \citep{naima} assuming pion decay and then integrated along the line of sight to produce Gammapy-compatible flux maps.\\
Each source observation was simulated with 200 hr exposure with the respective IRFs\footnote{https://zenodo.org/records/5499840} (ASTRI Mini-Array for northern sources; CTAO-South for southern sources) \citep{astriirf,ctairf}. Simulations accounted for astrophysical and instrumental backgrounds and were performed on an $8^\circ \times 8^\circ$ field of view.
In order to describe the morphology, we examined radial profiles of background-subtracted excess emission. To estimate the position for each source, we measured the peak of the emission position for TeV halos or the barycentre for YMSCs following area correction of the pixel and effective area, and the 90\% containment radius ($r_{90}$). We then computed and analysed the gamma-ray emission profiles from the centre of each source and modeled them to understand what morphological features allow a more secure classification. The profiles, normalized to their emission peak and rescaled to $0 < r/r_{90} < 1$, are shown in Figure \ref{fig:radial_profiles_spheric}. We used a modified Lorentzian function to fit the radial profiles:
\[f(r;  r_\textrm{p}, N, a, w) = \frac{N}{1 + \left(\frac{r-  r_\textrm{p}}{w}\right)^2} \, \left[1 + a \, (r-  r_\textrm{p})^2\right] \]
The fit parameters are the normalization $N$, the curve width $w$, the emission peak position $r_p$, and the anisotropy $a$. Figure \ref{fig:radial_profiles_spheric} illustrates how the $(r_p, a)$ parameter combination reliably separates YMSCs from TeV halos: YMSCs feature broader emission peaks ($r_p$) and lower anisotropy ($a$) and are therefore below the dash-dotted line, whereas TeV halos are in above the dash-dotted line because they are peak-centered and have higher anisotropy in their emission.

\section{LHAASO Sources Classification with IACTs}
\begin{figure}
\centering
\includegraphics[width=0.48\linewidth]{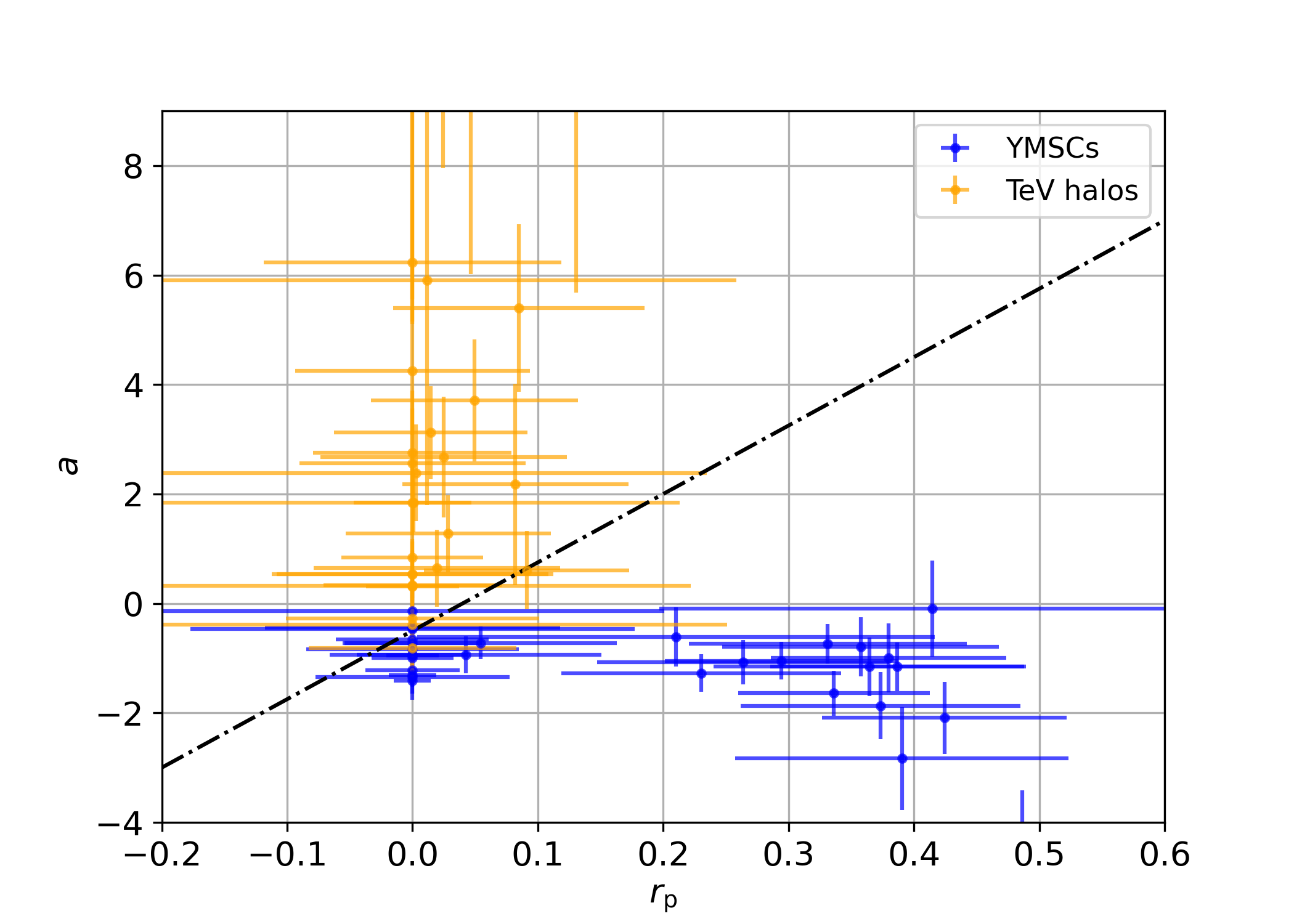}
\includegraphics[width=0.48\linewidth]{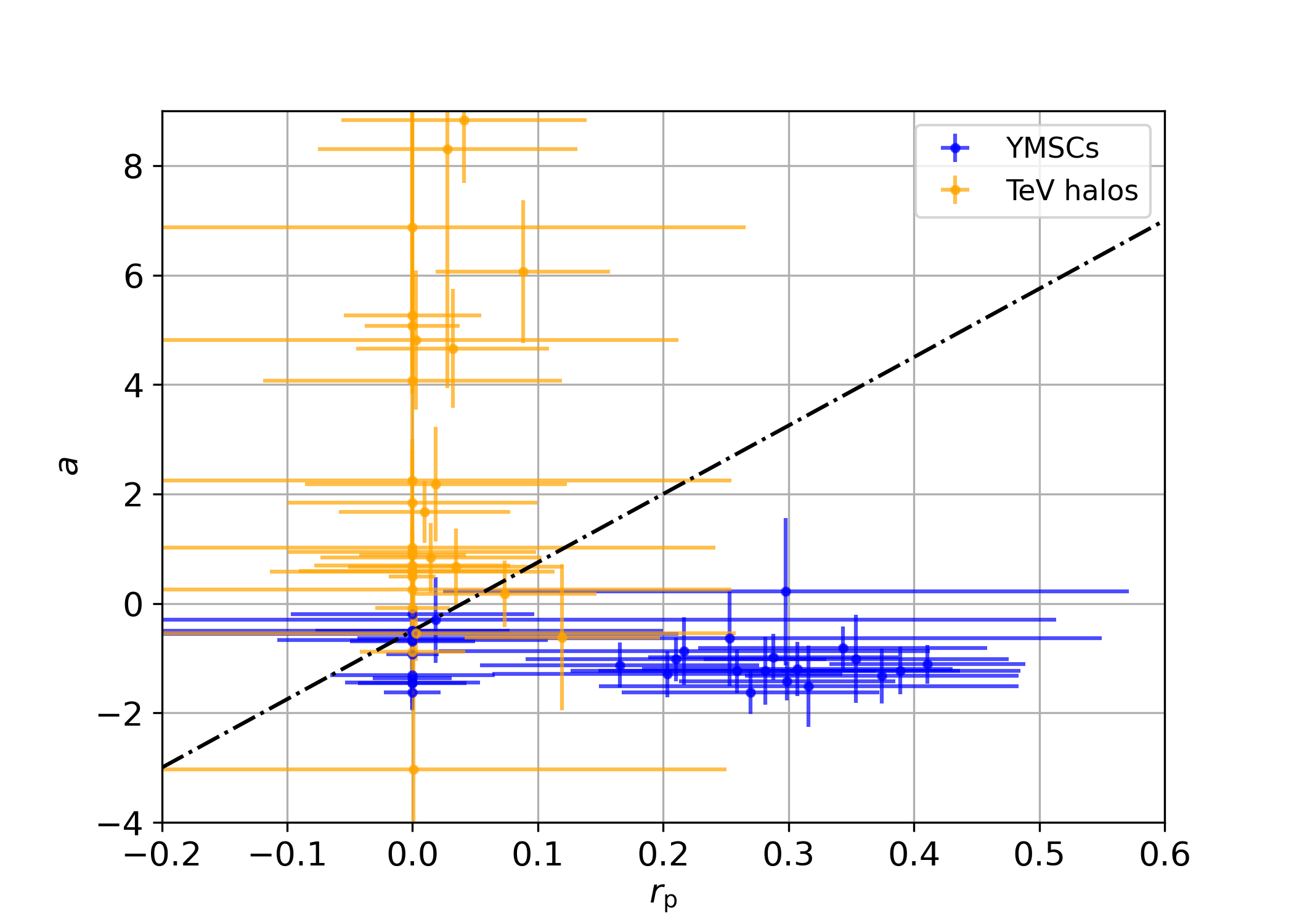}
\caption{\textit{Left:} Anisotropy and emission peak values (in units of $r_{90}$) for all LHAASO sources simulated for 50 hr using the CTAO IRF. Orange and blue crosses represent TeV halos and YMSCs, respectively. The black dot-dashed line is the reference from Figure~\ref{fig:radial_profiles_spheric}. \textit{Right:} Same as left, but for the case of the ASTRI Mini-Array IRF.}
\label{fig:scatter}
\end{figure}
In order to exercise the classification capabilities of CTAO and the ASTRI Mini-Array, we have applied our classification method to the unidentified extended sources in the first LHAASO catalog \citep{lhaaso_cat}. These are all the extended sources without a match in the TeVcat or 3HWC catalog within a searching radius of $r_{sr} =\sqrt{\sigma^2_{p,95} + r_{39}^2 + (0.3^\circ)^2}$, where $\sigma_{p,95}$ and $r_{39}$ are the position error and 39\% LHAASO containment radius of the source emission, both in degrees. We excluded sources with angular extensions $>4^\circ$, detected by either one or both LHAASO detectors. This gave us 29 WCDA and 26 KM2A sources, 22 of which were common for the two instruments.\\
Every source is parameterized in the LHAASO catalog with independent spectral (power-law) and spatial (2D Gaussian) components. We built 3D models for those with spatial widths $\sigma < 1.33^{\circ}$, assuming energy-independent morphology. Joint detections were modeled with the broken power law with break at 25~TeV, and single-instrument detections used the respective simple power law. We employed template-based spatial profiles out to $3\sigma$, with the profile of either YMSCs or TeV halos as shown in Section 3. Simulations were run with both the CTAO and ASTRI Mini-Array IRFs, for all sources. Observation pointings were fixed at the LHAASO-reported position, or, where there was a WCDA-KM2A joint detection, at the photon-weighted average position. We made an assessment of different exposure times (10, 50, 100, and 300 hours) to determine the minimum observation livetime required for successful source classification, using radial profiles corrected for instrumental effects. Every profile was fitted with the modified Lorentzian, allowing us to extract emission peaks and anisotropy parameters.\\
Based on the peak of emission alone ($\Delta r_p \geq 0.15\, r_{90}$), 6 (CTAO) and 3 (ASTRI Mini-Array) sources can be identified correctly with just 10 hours of exposure. This rises to 45.5\% (CTAO) and 40.9\% (ASTRI Mini-Array) in 50 hours, and up to 74.2\% and 69.7\% respectively with 300 hours. Adding the anisotropy parameter significantly improves the classification: with 50 hr, 68.2\% of the sources are correctly classified with CTAO IRFs, and 65.2\% with ASTRI, with some borderline cases due to overlapping uncertainties. Classification outcomes are given in Figure \ref{fig:scatter}.

\section{Conclusions}
In this work, we introduced new criteria to classify extended gamma-ray sources at energies $E \geq 1$~TeV based on their morphological features. Using simulations of five YMSCs and two TeV halos, we modeled their emission as observed by next-generation IACTs, specifically CTAO and the ASTRI Mini-Array. The radial excess profiles of these sources were characterized with an empirical function, highlighting the emission peak location and the anisotropy as the key discriminate parameters. The anisotropy parameter from the modified Lorentzian model proved particularly effective in distinguishing between YMSC- and TeV halo-like profiles.
Applying the classification to our test set of 7 simulated sources, we achieved 100\% accuracy in identifying their nature, both through the emission peak and anisotropy parameters. Each profile also consistently matched the expected classification region in the $r_p -a$ parameter space.\\
We extended our analysis to the unidentified extended sources in the first LHAASO catalog simulating their observation with CTAO and ASTRI Mini-Array IRFs. Our results show that a substantial fraction of the sources can be reliably classified with tens of hours of observation. Combining peak and anisotropy parameters, up to 68.2\% (CTAO) and 65.2\% (ASTRI) of the sources are correctly identified with just 50 hours of observation. Sources remaining ambiguous tend to have low surface brightness, small angular sizes, or both—limiting morphological classification.\\
Roughly 2200 Fermi sources remain unassociated, and an estimated 7\% could be related to embedded YMSCs \cite{peron}. At TeV energies, a large number of unidentified sources persists \cite{lhaaso_cat}, many of which may be attributed to massive stellar clusters. One of our objectives is to develop robust morphological and spectral identification criteria for YMSCs in gamma-ray surveys. When observational data from ASTRI Mini-Array and CTAO become available, the proposed classification method will support the identification of extended sources, improving our understanding of their nature and the underlying particle acceleration mechanisms in environments such as YMSCs.

\footnotesize
\paragraph*{Acknowledgements}
This work was produced while A.B. was attending the PhD program in in Space Science and Technology at the University of Trento, Cycle XXXVIII, with the support of a scholarship financed by the Ministerial Decree no. 351 of 9th April 2022, based on the NRRP - funded by the European Union - NextGenerationEU - Mission 4 "Education and Research", Component 1 "Enhancement of the offer of educational services: from nurseries to universities” - Investment 4.1 “Extension of the number of research doctorates and innovative doctorates for public administration and cultural heritage” - CUP C53C22000430006. 
A.B. acknowledges financial support from the European Union—Next Generation EU under the project IR0000012—CTA+ (CUP C53C22000430006), announcement N.3264 on 28/12/2021: “Rafforzamento e creazione di IR nell’ambito del Piano Nazionale di Ripresa e Resilienza (PNRR)”.

\normalsize

\bibliographystyle{aa}
\bibliography{references}

\end{document}